\documentclass[aps,pra,prl,nopacs,twocolumn,twoside,floatfix,a4,superscriptaddress]{revtex4}
\usepackage{hyperref}
\usepackage{ulem}
\usepackage{amsmath, amssymb,amsthm}
\usepackage{color}
\usepackage{graphicx,epsfig}
\usepackage{times}
\usepackage{tikz}
\theoremstyle{plain}
\newtheorem{theorem}{Theorem}
\newtheorem*{cor}{Corollary}
\theoremstyle{definition}
\newtheorem{definition}{Definition}
\newtheorem{lemma}{Lemma}
\newcommand{\defeq}{\mathrel{:\mkern-0.25mu=}}

\definecolor{matty}{rgb}{0,0,1}
\definecolor{gnz}{rgb}{0,0,1}

\def\x{\textbf{x}}
\def\a{\textbf{a}}

\begin{document}

\title{Maximally nonlocal theories cannot be maximally random}


\author{Gonzalo de la Torre}
\email{gonzalo.delatorre@icfo.es}
\affiliation{ICFO-Institut de Ciencies Fotoniques, Mediterranean Technology Park,
08860 Castelldefels (Barcelona), Spain}
\author{Matty J. Hoban}
\email{matthew.hoban@icfo.es}
\affiliation{ICFO-Institut de Ciencies Fotoniques, Mediterranean Technology Park,
08860 Castelldefels (Barcelona), Spain}
\author{Chirag Dhara}
\affiliation{ICFO-Institut de Ciencies Fotoniques, Mediterranean Technology Park,
08860 Castelldefels (Barcelona), Spain}
\author{Giuseppe Prettico}
\affiliation{ICFO-Institut de Ciencies Fotoniques, Mediterranean Technology Park,
08860 Castelldefels (Barcelona), Spain}
\author{Antonio Ac\'{i}n}
\affiliation{ICFO-Institut de Ciencies Fotoniques, Mediterranean Technology Park,
08860 Castelldefels (Barcelona), Spain}
\affiliation{ICREA--Institucio Catalana de Recerca i Estudis
Avan\c{c}ats, E--08010 Barcelona, Spain}


\begin{abstract}

Correlations that violate a Bell Inequality are said to be
nonlocal, i.e. they do not admit a local and deterministic
explanation. Great effort has been devoted to study how the amount
of nonlocality (as measured by a Bell inequality violation)
serves to quantify the amount of randomness present in observed
correlations. In this work we reverse this research program and
ask what do the randomness certification capabilities of a theory
tell us about the nonlocality of that theory. We find that,
contrary to initial intuition, maximal randomness certification cannot occur in maximally nonlocal theories. We go on and show that
quantum theory, in contrast, permits certification of maximal
randomness in all dichotomic scenarios. We hence pose the question
of whether quantum theory is optimal for randomness, i.e. is it
the most nonlocal theory that allows maximal randomness
certification? We answer this question in the negative by
identifying a larger-than-quantum set of correlations capable of
this feat. Not only are these results relevant to understanding
quantum mechanics' fundamental features, but also put fundamental
restrictions on device-independent protocols based on the
no-signaling principle.

\end{abstract}

\maketitle


From a physical perspective, all classical physics is deterministic and any apparent randomness is due to ignorance therefore not exhibiting \textit{intrinsic} randomness. Quantum theory is open to such a  possibility since it is a fundamentally probabilistic theory. However, since the early days of quantum theory, its seemingly `intrinsic'
unpredictability has been heavily debated even by some of its founding fathers
\cite{Einstein1935,Bohr1935}. 
A great advance came when John Bell
\cite{bell} identified limitations on any theory founded
on the following two basic physical principles: impossibility of
instantaneous signaling between distant locations (no-signaling principle); and
the existence of a complete set of variables of a system which, if known, would allow for deterministic predictions. Correlations among a number of distant parties that satisfy both
principles are called 'local' and are constrained by the now-eponymous Bell inequalities. Thus Bell established a fundamental link between the unpredictability of quantum mechanics with the concept of nonlocality \cite{bell}. In particular, assuming the validity of the no-signalling principle, a violation of a Bell inequality implies and certifies intrinsic randomness.

Recently, this deep connection between nonlocality and randomness has been made quantitative and exploited for information processing tasks \cite{bellnonloc,scarani}. Nonlocality-certified randomness represents an information resource in  the now well-established area of ``device-independent quantum information processing" \cite{randexp,colbeck,Colbeck2012,gallego,diqip,bhk}. In a device-independent protocol, no assumption is made about the inner-workings of the devices used and are thus regarded as black boxes. There is however a crucial assumption to every protocol and that is the assumption of the background theory dictating the devices' behaviour, e.g. whether the devices are quantum mechanical \cite{diqip}, or just compatible with no-signaling principle \cite{bhk}.

The assumption about the background theory is \textit{vital} given that quantum mechanics is not the most nonlocal theory respecting the no-signaling principle \cite{pr} and therefore capable of producing intrinsic randomness. Theories allowing for all nonlocal correlations \textit{only} restricted by the no-signaling principle are termed ``maximally nonlocal" since they produce the most nonlocality that a non-signalling theory can produce. Given the eminent role of nonlocality for randomness certification, the first intuition is to expect maximally nonlocal theories to have more powerful randomness certification capabilities than other theories.
Indeed, there are occasions where maximally nonlocal theories can certify randomness and quantum mechanics cannot even certify any randomness at all \cite{gyni}. On the other hand, for the Clauser-Horne-Shimony-Holt (CHSH) inequality \cite{chsh}, we can certify more randomness assuming only quantum mechanics (however not the maximal amount possible) rather than allowing maximally nonlocal correlations \cite{acinmassarpironio}. 
%
%

The main goal of our work is to understand the relationship between the
nonlocality and randomness of a theory and, in particular, what the randomness
capabilities of a theory tell us about the nonlocality allowed
within that theory. Notice that we do not consider misalignments, preparation errors or detection efficiencies as those sources of randomness are not fundamental to the physical theory and can, in principle, be reduced below any finite threshold. We are hence interested in studying only the fundamental differences in randomness certification between theories.


The first result is to show that the previous intuition is wrong: were the set of achievable physical correlations not more restricted than what the no-signaling principle allows, maximal randomness could not be certified in
any possible scenario irrespective of the number of parties, measurements or outcomes, i.e. maximally nonlocal theories cannot be maximally random.

Secondly, we focus on quantum theory and provide, in contrast, scenarios with an arbitrary number of parties where maximal randomness \textit{can} be certified. This should be compared with other works that showed that if maximally nonlocal theories were permitted in Nature we would have unimaginable computational and communicating power \cite{infc,trivial}. Here, being in a maximally nonlocal world \textit{limits} our information processing capabilities. This observation leads us to ask if the nonlocality of quantum theory is in some sense optimal for randomness certification. That is, is quantum theory the most nonlocal theory capable of certifying maximal randomness? Our final result answers this question in the negative: we identify a set of correlations larger than the quantum set that also permits the certification of maximal randomness.

%
%
%

\textit{Boxes and Bell tests}---We use the scenario of a Bell test to study the correlations observed among space-like separated measurements on systems within different physical theories. There are $N$ distant parties and each party makes a choice of measurement upon their system. These processes are arranged so that they define space-like separated events. The $N$ users have no knowledge of how a system or its measurement apparatus are prepared, they can only make measurement choices and observe classical outcomes. We even allow the possibility that a malicious agent prepared the devices and holds information about how they prepared their systems. We then model these parties as black boxes with the measurement choice for the $j$th party (for $j\in\{1,2,...,N\}$) being an input $x_j\in\{0,1,..,(M-1)\}$ where there are $M$ possible choices; the measurement outcome for this $j$th party is then the output $a_j\in\{0,1,...,(d-1)\}$ where for every party there are $d$ possible outcomes to the every measurement. Therefore, a string of ``dits" (generalization of bits to $d$ values) is produced in each round of a Bell test . The Bell test is then labelled by the parameters $(N,M,d)$. 

After a suitable number of uses of the boxes, the conditional probabilities $p_{\rm obs} (\textbf{a}|\textbf{x})$ for all values of $\textbf{a}=(a_1,\ldots,a_N)$ and $\textbf{x}=(x_1,\ldots x_N)$ that describe the observed process are obtained. These conditional probabilities form a full distribution $P_{\textrm{obs}}$ with elements $p_{\rm obs} (\textbf{a}|\textbf{x})$. In general, we will use an upper-case $P$ for a distribution and lower-case $p$ for an element of that distribution.

As mentioned, we must make an assumption on the \textit{theory} that governs the workings of our boxes. This has the effect of indicating whether our observed correlations $P_{\textrm{obs}}$ belong to a particular set of possible correlations. For example, if we say that our observed correlations result from a classical system or quantum system then the distribution $P_{\textrm{obs}}$ belongs to the set $C$ or $Q$ of correlations resulting from all possible classical and quantum systems respectively. We can also define the set $NS$ of all maximally nonlocal correlations. 

The set of quantum correlations $Q$ is contained in $NS$. 
This does not imply that the latter is as random as the former. The important distinction is that by dictating which theory is permitted, we bound the power of the malicious agent that can, in principle, prepare our devices. Therefore, upon obtaining our observed statistics $P_{\textrm{obs}}$, if we vary the theory that describes the source of these correlations then we allow a malicious agent to prepare the devices using different (even supra-quantum) resources. The agent can then use this knowledge of the preparation to improve their predictive power thus leading to different implications for randomness certification.

Every set is convex because we can always prepare a convex mixture (by tossing a biased coin) of systems thus giving a convex mixture of correlations resulting from each system. So, in addition to the Bell scenario dictated by $(N,M,d)$ we stipulate the set of correlations $T$ to which our observed correlations can belong.

Every convex set can be described in terms of its extreme points. In the case of probability distributions $P$, the extreme points are those distributions that cannot be expressed as a convex combination of other distributions in the set. 
An immediate corollary of this property of convex sets is that the observed correlations $P_{\textrm{obs}}$ can be written as a convex combination of the extreme points of a set $T$: therefore $P_{\textrm{obs}}=\sum_{\textrm{ext}}q_\textrm{ext}P_{\textrm{ext}}$ where $P_{\textrm{ext}}$ is an extreme point of the set $T$ and $q_\textrm{ext}\geq 0$ is a probability distribution over these points such that $\sum_{\textrm{ext}}q_\textrm{ext}=1$.

\textit{Randomness Certification}---In randomness certification we take the standard approach of using just one particular $\textbf{x}_{0}$ from which to obtain a dit-string of length $N$ that is hopefully random. The (potentially malicious) provider of the boxes knows this string $\textbf{x}_0$ in advance however does not know \textit{when} this string is input into the boxes. Correlations for the rest of the inputs $\{\textbf{x}|\textbf{x}\neq\textbf{x}_{0}\}$, encapsulated by $P_{\textrm{obs}}$, are used to certify that the randomness obtained from $\textbf{x}_{0}$ is intrinsically random. To measure the randomness of the outputs obtained from input $\textbf{x}_{0}$ given observed correlations $P_{\textrm{obs}}$, we require the \textit{guessing probability} $G^{T}(\textbf{x}_{0},P_{\textrm{obs}})$: the probability for a malicious agent to predict the most likely outcome for input $\textbf{x}_{0}$ given that the agent has complete knowledge of how a box is prepared within a theory with a corresponding set of correlations $T$. The larger this guessing probability the less random are the outputs and thus we have a measure of randomness. 

Since the user of the box has no knowledge of how it was prepared, we must assume that all possible ways of producing $P_{\textrm{obs}}$ from extreme points $P_{\textrm{ext}}$ can be utilized such that $P_{\textrm{obs}}=\sum_{\textrm{ext}}q_\textrm{ext}P_{\textrm{ext}}$ and the malicious agent knows $q_\textrm{ext}$ perfectly. Indeed, the agent may know what is the most advantageous distribution $q_\textrm{ext}$ to maximize his chances of guessing the output. On the other hand, for each of these extreme points $P_{\textrm{ext}}$ we can evaluate the guessing probability easily since there is a unique way of preparing this probability distribution (from the set of correlations $T$). Therefore, the guessing probability for extreme points is $G^{T}(\textbf{x}_{0},P_{\textrm{ext}})=\textrm{max}_{\textbf{a}}\textrm{ }p_{\textrm{ext}}(\textbf{a}|\textbf{x}_{0})$ where $p_{\textrm{ext}}(\textbf{a}|\textbf{x}_{0})$ is an element of $P_{\textrm{ext}}$. Collating all of this information, we obtain the following optimization:
\begin{align}
& G^T(\x_0,P_{\textrm{obs}})= \max_{\{q_\textrm{ext},P_{\textrm{ext}}\} } \sum_{\textrm{ext}} q_\textrm{ext} G^{T}(\textbf{x}_{0},P_{\textrm{ext}}) \nonumber \\
 &\text{subject to:} \nonumber \\
&P_{\textrm{obs}}=\sum_{\textrm{ext}}q_\textrm{ext}P_{\textrm{ext}},P_{\textrm{ext}}\in T. \label{sumupto} 
\end{align}
Immediately we see that for classical correlations (when $T=C$) $G^{C}(\x_0,P_{\textrm{obs}})=1$ since $\textrm{max}_{\textbf{a}}\textrm{ }p_{\textrm{ext}}(\textbf{a}|\textbf{x}_{0})=1$ for all $P_{\textrm{ext}}$. This follows from the well-known fact that any classical correlations can be decomposed as a mixture of deterministic points. This highlights the need for non-classical, or nonlocal correlations for randomness certification.

%
%

It is worth noting that we are explicitly assuming the independence between the preparation components labelled by $P_{\textrm{exp}}$ and the measurement settings $\textbf{x}$. This is commonly known as the \textit{freedom of choice} assumption. Recent work has shown that this assumption can even be relaxed by implementing randomness amplification protocols \cite{Colbeck2012,gallego,minentropy}.

In what follows we will perform the optimization in \eqref{sumupto} for different sets of correlations, in particular maximally nonlocal and quantum correlations labelled $NS$ and $Q$ respectively. In particular we ask the question of whether a theory with correlations $T$ can certify maximal randomness which exactly means if for any observed correlations $P_{\textrm{obs}}\in T$ in any scenario $(N,M,d)$, we can obtain $G^T(\x_0,P_{\textrm{obs}})=\frac{1}{d^{N}}$. 

\textit{Maximally nonlocal correlations}---The set $NS$ of maximally nonlocal correlations is the set of multipartite correlations solely restricted by the no-signaling principle. Here we permit any valid normalized probability distribution $P$ with all elements satisfying $1\geq p(\textbf{a}|\textbf{x})\geq 0$ where marginals are well-defined. That is, the probabilities (correlations) satisfy $\sum_{\textbf{a}}p(\textbf{a}|\textbf{x})=1$. To prevent instantaneous signaling it is important that
\begin{align}
\sum_{a_k} p(a_1,\ldots,a_k\,\ldots a_N|x_1,\ldots,x_k,\ldots,x_N)
\label{ns}
\end{align}
is independent of $x_k$ for all $k$.

Now that this set is defined we present our first result.
\newline

\noindent
\textbf{Result 1}: \textit{Maximally nonlocal theories can never be maximally random.}

Were the physically achievable correlations solely restricted by the no-signaling principle, the maximum amount of certifiable randomness in an arbitrary Bell scenario $(N,M,d)$ would be bounded through the intrinsic predictability by
\begin{equation}
G^{NS}(\x_0,P_{\textrm{obs}})\geq\frac{1}{d^N-(d-1)^N},
\end{equation} 
for any probability distribution $P_{\rm obs}\in NS$ and all inputs $\textbf x_0$.
\newline

To prove this result we only need to consider the randomness of the extreme points $P_{\textrm{ext}}$ of $NS$ as indicated by \eqref{sumupto}. Our proof is based on the simple observation that if for a particular $\textbf{x}_{0}$ of correlations $p(\textbf{a}|\textbf{x})$, $n$ values 
 are equal to zero then $\max_{\textbf{a}}p(\textbf{a}|\textbf{x}_0)\geq \frac{1}{d^{N}-n}$. Result 1 then follows from Theorem \ref{thmone} in the Appendix, which proves that some given non-signalling correlations $p(\textbf{a}|\textbf{x})$ cannot be extreme if there exists a string of inputs $\textbf{x}_{0}$ such that the number of terms $p(\textbf{a}|\textbf{x}_0)$ that are equal to zero is smaller than $(d-1)^N$.

It is worth mentioning two facts. First, this result indicates an important limitation on maximally nonlocal theories. In fact, the gap between the ideal maximal randomness and that achievable in maximally nonlocal theories is unbounded. Second, the derived bound is, in general, not tight. For instance, all extreme non-signaling correlations in Bell test scenarios $(2,M,2)$ were obtained in \cite{masanes,bp} and in this case $G^{NS}(\x_0,P_{\textrm{obs}})\geq1/2$ whereas our bound gives $1/3$. Interestingly, the same difference appears in the $(3,2,2)$ scenario: looking at all the extreme points,  classified in \cite{bancal}, the maximal randomness is equal to $1/6$, while our bound predicts $1/7$. However, in the asymptotic limit of $d\rightarrow\infty$ our bound gives $\frac{1}{O(d^{N-1})}$, which can be shown to be tight by comparing it with the results in \cite{aolita}. We now move to randomness certification in quantum theory.

\textit{Quantum Correlations}---Let $\rho\geq 0$ be some quantum state and $O_{a_j}^{x_j}$ be some measurement operators (technically a positive operator valued measure, POVM) for input $x_j$ and output $a_j$. We say a probability distribution $P_{\rm obs}\in Q$ belongs to the quantum set of correlations if it can be written as $p_{\rm obs}(\textbf{a}|\textbf{x})=\textrm{tr}(\rho \bigotimes_{j=1}^{N}O_{a_j}^{x_j})$.

Characterizing the set of correlations achievable in this way is a great open problem in quantum information theory. Therefore, in what follows, rather than solving exactly the optimization problem (\ref{sumupto}), we consider a relaxation that provides a lower bound to the intrinsic randomness. Instead of considering all convex combinations of extreme points of $Q$ that reproduce the observed statistics, we ask for convex combination of extreme points that give an observed violation of a Bell inequality. Given that a Bell inequality is just a linear combination of probabilities $p(\textbf{a}|\textbf{x})$ over all inputs $\textbf{a}$ and outputs $\textbf{x}$, let us define the following inner product between correlations $P_{\rm obs}$ and Bell inequality $B$ that computes the Bell violation $B\cdot P_{\textrm{obs}}\equiv\sum_{\textbf{a},\textbf{x}} \beta_{\textbf{a},\textbf{x}} p_{\textrm{obs}} (\textbf{a}|\textbf{x})=q_{\textrm{obs}}$, where the real coefficients $\beta_{\textbf{a},\textbf{x}}$ define the Bell inequality $B$.

Computing a lower bound to the intrinsic predictability $G^{Q}(\textbf{x}_{0},P_{\rm obs})$, certified this time by an observed violation of a Bell inequality, then amounts to solving the following optimization problem, a relaxation of \eqref{sumupto}:
\begin{align}
& G^Q(\x_0,P_{\textrm{obs}})\leq\max_{\{q_\textrm{ext},P_{\textrm{ext}}\} } \sum_{\textrm{ext}} q_\textrm{ext} G^{T}(\textbf{x}_{0},P_{\textrm{ext}}) \nonumber \\
 &\text{subject to:} \nonumber \\
&\sum_{\textrm{ext}}q_\textrm{ext}(B\cdot P_{\textrm{ext}})=q_{\textrm{obs}},P_{\textrm{ext}}\in Q.
\end{align}
Since we are interested in the maximal amount of randomness allowed by quantum mechanics, we will restrict our study to maximal quantum violation of a Bell Inequality $q_{\rm obs}\equiv q_{\rm max}$. In \cite{dhara} a method was provided to detect when the maximal quantum violation of a Bell inequality certifies that the outputs are maximally random. The method has the advantage that it can be easily applied, but unfortunately it only works under the assumption that the maximal quantum violation of the inequality is unique. The uniqueness of the maximal quantum violation is in general hard to prove. However, in what follows, we consider Bell inequalities for which the uniqueness of the maximal violation can be proven using the results of Refs. \cite{franz,millershi}. This then allows us to apply the simple method in \cite{dhara} and prove the following result.
\newline

\noindent
\textbf{Result 2}: \textit{Quantum theory is maximally random in all dichotomic scenarios.}

Assuming the set of physically achievable correlations to be the quantum set, the maximum amount of certifiable randomness in the family of Bell test scenarios $(N,M,2)$ is maximal: $G^{Q}(\textbf{x}_{0},P_{\textrm{obs}})=\frac{1}{2^{N}}$.
\newline

We prove this result in the Appendix by generalizing the results of \cite{dhara} to all $N$ via a Bell inequality introduced in \cite{nmbqc}. We actually prove Result 2 for the $(N,2,2)$ scenario but this trivially applies to the $(N,M,2)$ since we can always ignore $(M-2)$ of the inputs for each party. While our proof does not apply to the case of two parties, it has been shown in \cite{nlvsran} that for the $(2,2,2)$ scenario an amount of randomness arbitrarily close to the maximum of $2$ random bits can be certified in some limit. Additionally, numerical and analytical evidence indicates that exactly $2$ bits of maximal randomness can be attained in the $(2,3,2)$ scenario \cite{dhara}. All of this serves to show that quantum correlations certify maximal randomness even if maximally nonlocal theories can never do this.

We have shown the difference for randomness certification of two sets of correlations; the maximally nonlocal set and the quantum set. 
A natural question is whether this contrast highlights the \textit{uniqueness} of quantum correlations. Just as various information theoretic principles aim to highlight single-out quantum theory \cite{infc,trivial,locorth}, is $Q$ the only set capable of certifying maximal randomness? We now address this question.

\textit{Supra-quantum Correlations}---Navascu\'{e}s, Pironio and Ac\'{i}n introduced a means to approximate the set of quantum correlations which was an infinite hierarchy of semi-definite programs \cite{npa}. 
For example, the first non-trivial level of this hierarchy is $Q^1$ and this set is provably larger than the set of quantum correlations $Q$ \cite{nav}. Already in the work of Pironio et al in Ref. \cite{randexp} these first few levels in the hierarchy were used to lower bound the amount of randomness certified for quantum correlations. 
In the Appendix, we introduce a modification to the set $Q^1$ in the tripartite setting called $Q^{1+ABC}$ that is strictly larger than the quantum set. On the other hand, this set also allows for maximal randomness certification. This represents the third main result of this work.
\newline

\noindent
\textbf{Result 3}: \textit{There exist post-quantum theories that can also certify maximal randomness.}

Were the physically achievable correlations those of the strictly larger than quantum set $Q^{1+ABC}$, maximal randomness could also be certified in the Bell test scenario $(3,M,2)$ i.e. $G^{Q^{1+ABC}}(\textbf{x}_{0},P_{\textrm{obs}})=\frac{1}{8}$.
\newline

The proof of this result is presented in the Appendix. The crucial element in this proof is showing that there is only one probability distribution in the set $Q^{1+ABC}$ that maximally violates the Mermin inequality \cite{Mermin1990a} allowing us to use the results in Ref. \cite{dhara}.

At first, this result may seem disappointing but there are other examples of limitations to recovering quantum correlations from information principles. For example it is known that we need truly multipartite principles \cite{multi}. It has also been shown that other information principles will never recover quantum mechanical correlations \cite{nav} and our work fits squarely within this foundational research program.

\textit{Discussion}---We have shown that correlations in maximally nonlocal theories and quantum theory have drastically different consequences for randomness certification. Therefore, if we assume Nature does not abide by a nonlocality-restricted theory such as quantum theory it could severely limit its randomness capabilities. One can see this as a result of maximally nonlocal correlations having correlations between the outputs \textit{for all} inputs, but quantum theory cannot produce such strong correlations.
Let us illustrate this point by revisiting the CHSH scenario of $(2,2,2)$. Here all extremal correlations in maximally nonlocal theories are equivalent to the so-called Popescu-Rorhlich (PR) box \cite{pr}. This box always fulfils the condition: $x_1\cdot x_2=a_1\oplus a_2~{\rm mod~2}$ \footnote{This condition can be seen as a rewriting of the Clauser-Horne-Shimony-Holt (CHSH) inequality \cite{chsh}, sometimes referred to as the CHSH game.}. Therefore, knowing the inputs and one of the outputs, we can \textit{perfectly determine} the other output. Quantum correlations, however, cannot produce these perfect correlations thus introducing more randomness.

These results are not only of foundational interest but have application in randomness extraction, certification and amplification. For example, in Ref. \cite{randexp} a lower bound on certifiable randomness was obtained using only the no-signaling principle, and this bound has found applications in other protocols (e.g. Ref. \cite{vv}). 
An interesting follow-up question is to determine the \textit{exact} maximum randomness allowed just by the no-signalling principle, a fundamental number providing a quantitative link between randomness and no-signalling.

We also showed that certain supra-quantum correlations can also exhibit maximal randomness. 
This last result indicates that quantum theory is not so special from an information theoretic perspective (cf. Ref. \cite{nav}). 
The set of quantum correlations is notoriously difficult to define but maximally nonlocal theories have a simple description. The fact that there exists a set of correlations that has a relatively simple description but facilitates maximal randomness certification provides a ``third way" for the design and analysis of future protocols.

\textit{Acknowledgements}---We acknowledge financial support from the ERC Consolidator Grant QITBOX, the John Templeton Foundation, an SGR from the Generalitat de Catalunya, and the Spanish MINECO, through an FPI grant and projects Intrinqra and Chist-Era DIQIP.

\onecolumngrid
\appendix

\section{Appendix}

\subsection{Proof of Result 1}

In this section we show that it is impossible for maximally nonlocal theories to produce maximal randomness, or more specifically Result 1 in the main text. Result 1 follows from the following Theorem, which provides a bound on the number of non-zero entries in extreme non-signaling correlations.
\newline

\begin{theorem} Let $p(\textbf{a}|\textbf{x})$ be an extreme probability distribution in the set $NS$ in an arbitrary 
Bell test scenario $(N,M,d)$. For a given combination of settings $\textbf{x}_0$, 
denote by $n(\textbf{x}_0)$ the number of probabilities 
$p(\textbf{a}|\textbf{x}_{0})$ 
that are equal to zero and define $n=\min_{\textbf{x}_0} n(\textbf x_0)$. 
Then, $n\geq (d-1)^N$.\label{thmone}
\end{theorem}

\textit{Proof}: The proof of the result follows from a relatively simple counting argument. First we introduce some useful notation to describe the marginals of a probability distribution. If we have a distribution $P$ with elements $p(\a|\x)$ and we have a set $\mathcal{J}\subseteq\{1,2,...,N\}$ of the $N$ parties then the probability distribution only over these parties in $\mathcal{J}$ is $p(\a^{\mathcal{J}}|\x^{\mathcal{J}})=\sum_{a_{j}|j\notin\mathcal{J}}^{N}p(\a|\x)$ where $\a^{\mathcal{J}}$ and $\x^{\mathcal{J}}$ are $\a$ and $\x$ consisting only of elements $a_{j}$ and $x_{j}$ respectively for all $j\in\mathcal{J}$. Following a simple generalization of Ref. \cite{collins}, a probability distribution (for any input $\textbf{x}_{0}$) satisfying the no-signalling principle can be parametrized by $p(\a^{\mathcal{J}}|\x^{\mathcal{J}})$ for all possible sets $\mathcal{J}$. What is more, due to normalization we only consider $(d-1)$ outputs for each party in all of these distributions. Therefore the probability $p(\a|\x)$ is a function of $p(\a^{\mathcal{J}}|\x^{\mathcal{J}})$ for all $\mathcal{J}$ but the elements $a_{j}$ of $\a^{\mathcal{J}}$ only range over $(d-1)$ values. Apart from when $\mathcal{J}$ contains all $N$ parties, every other marginal probability $p(\a^{\mathcal{J}}|\x^{\mathcal{J}})$ will result from another probability distribution $p(\textbf{a}|\textbf{x}')$ for $\textbf{x}'\neq \textbf{x}_{0}$ by summing over outputs of the appropriate parties. Therefore the values of these marginals are fixed by probabilities for inputs $\textbf{x}'\neq\textbf{x}_{0}$ and the only free parameters defining $p(\textbf{a}|\textbf{x}_{0})$ are the $(d-1)^{N}$ probabilities when $\mathcal{J}$ contains all $N$ parties.

Clearly, the space of this $d^N$-outcome probability distributions $p(\textbf{a}|\textbf{x}_{0})$ is convex. Moreover, if $p(\textbf a|\textbf x_0)$ is not an extreme point in this space, neither are the original correlations $p(\textbf a|\textbf x)$ in the original non-signalling space. As mentioned, when restricted to the specific setting $\textbf{x}_0$, there are $(d-1)^N$ free parameters. Now, the hyperplanes defining this convex space correspond to the positivity constraints defined by the $d^N$ probabilities $p(\textbf a|\textbf x_0)$. An extreme point in this space of dimension $(d-1)^N$ should then be defined by the intersection of $(d-1)^N$ hyperplanes. This implies that a necessary condition for the correlations $p(\textbf a|\textbf x)$ to be extreme is that at least $(d-1)^N$ probabilities $p(\textbf a|\textbf x_0)$ are zero for each value of $\textbf x_0$. This completes the proof. $\square$

\subsection{Proof of Result 2}

In this section we show that it is possible to obtain $N$ bits of global randomness for all $N$. In Ref. \cite{dhara} it was shown how to achieve and certify $N$ bits of global randomness for all odd $N$. Here, we show that there is a Bell inequality in the $(N,2,2)$ setting for all $N$ (which is a generalization of the Mermin inequality first studied in \cite{nmbqc}) which if maximally violated, gives $N$ bits of global randomness. We use the tools developed in Ref. \cite{dhara} to obtain maximal randomness based on the symmetries of the inequality we use.

First, we need to introduce some notation. As standard in the literature, we introduce the $n$-party correlators where $n\leq N$. Take a subset $J_{n}\subseteq\{1,2,...,N\}$ of $n$ parties from all $N$ parties. Then associated with this subset and a string of inputs $x_{J_{n}}=(x_{j},x_{j'},...,x_{k})$ and a string of outputs $a_{J_{n}}=(a_{j},a_{j'},...,a_{k})$ where $j$, $j'$, $k\in J_{n}$ and a marginal probability distribution $p(a_{J_{n}}|x_{J_{n}})$. We then define the correlators to be
\begin{equation}
\langle x_{J_{n}}\rangle\defeq 2\left(\sum_{a_{J_{n}}}\alpha p(a_{J_{n}}|x_{J_{n}})\right)-1,
\end{equation}
with $\alpha=1+\sum_{k\in J_{n}}a_{k}\textrm{ mod } 2$. We can define the full joint probabilities in terms of these correlators as
\begin{equation}\label{distro}
p(\textbf{a}|\textbf{x})=\frac{1}{2^{N}}\sum_{J_{n}}(-1)^{\sum_{k\in J_{n}}a_{k}}\langle x_{J_{n}}\rangle,
\end{equation}
where we take a sum over all $2^{N}$ subsets of $N$ parties (including the empty set).

The Bell inequality that will concern is the following inequality discussed in Ref. \cite{nmbqc}:
\begin{equation}\label{ineq1}
\sum_{\textbf{x}}(-1)^{f(\textbf{x})}\delta^{g(\textbf{x})}_{0}\langle x_{J_{n}}\rangle\leq \epsilon <2^{N-1},
\end{equation}
where $f(\textbf{x})=\sum_{j=1}^{N-1}x_{j}\left(\sum_{k=j+1}^{N}x_{k}\right)\textrm{ mod } 2$ and $g(\textbf{x})=\sum_{j=1}^{N}x_{j}\textrm{ mod } 2$. As indicated the upper-bound for local hidden variables $\epsilon$ is strictly less than $2^{N-1}$, the number of terms in the sum. Crucially, quantum mechanics can violate this inequality and achieve the algebraic upper bound of $2^{N-1}$ as shown in Ref. \cite{nmbqc} using a Greenberger-Horne-Zeilinger state. Also, there is only one probability distribution that maximally violates this inequality as can be shown by applying the techniques of Ref. \cite{franz} or by a self-testing argument due to \cite{millershi}. In Ref. \cite{dhara} this property of \textit{uniqueness} of a probability distribution maximally violating an inequality was used to prove that global randomness can be generated from a Bell test.

We need to use an input $\textbf{x}'$ that does not appear in the left-hand-side of \eqref{ineq1}, since the outputs of measurements for inputs in \eqref{ineq1} will be highly correlated, and thus not random. We need to show that for input $\textbf{x}'$, the probability $p(\textbf{a}|\textbf{x}')$ in \eqref{distro} is equal to $\frac{1}{2^{N}}$ for all $\textbf{a}$. This occurs if all correlators satisfy $\langle x_{J_{n}}\rangle=0$ for all (non-empty) subsets $J_{n}$ and $\langle x_{J_{n}}\rangle=1$ for $J_{n}=\emptyset$, the empty set when $n=0$. The aim of this section is to show that this is true.

To do this, we utilize the tools in Ref. \cite{dhara} where we perform transformations on the data obtained in a Bell test that do not affect the correlators that appear in the Bell inequality of \eqref{ineq1}. These transformations affect correlators that do appear in the inequality. If we take the unique probability distribution that maximally violates \eqref{ineq1} then, under these transformations, it still violates the same inequality maximally. If we call the original probability distribution $P$ with elements $p(\textbf{a}|\textbf{x})$ and the transformed distribution $P'$, then $P=P'$, and so all correlators resulting from these two distributions must be equal as well. If one of these symmetry transformations is to flip an outcome of a measurement depending on the choice of input then this can alter correlators, e.g. $a_{1}\rightarrow a_{1}\oplus x_{1}$, then all correlators $\langle x_{J_{n}}\rangle$ that contain $x_{1}=1$ have their sign flipped as $\alpha=1+\sum_{k\in J_{n}}a_{k}\textrm{ mod } 2\rightarrow 2+\sum_{k\in J_{n}}a_{k}\textrm{ mod } 2$. However, due to uniqueness of quantum violation this implies that the correlators before and after the transformation are equal, so in the case that the transformation flips the sign of the correlator then $\langle x_{J_{n}}\rangle=-\langle x_{J_{n}}\rangle=0$. This thus demonstrates a way to show that correlators are zero for particular distributions.

For clarity we introduce the notation to show when a correlator's sign is flipped. Our symmetry operations are captured by an $n$-length bit-string $\textbf{s}$, where if the $j$th element $s_j$ is zero, then we flip $a_{j}$ for the choice of input $x_{j}=0$, and if $s_{j}=1$, then we flip $a_{j}$ for choice of input $x_{j}=1$. Then the correlator $\langle x_{J_{n}}\rangle$ under the symmetry transformation described by $s$ is mapped to $(-1)^{N-H(\textbf{x},\textbf{s})}\langle x_{J_{n}}\rangle$ where $H(\textbf{x},\textbf{s})$ is the Hamming distance between the bit-strings $\textbf{s}$ and $\textbf{x}$: the number of times $x_{j}\neq s_{j}$ for bit-strings $\textbf{x}$, $\textbf{s}$. Another way of writing the Hamming distance is
\begin{equation}\label{hamming}
H(\textbf{x},\textbf{s})=\sum_{j=1}^{N}x_{j}+s_{j}\textrm{ mod } 2=\sum_{j=1}^{N}x_{j}+s_{j}-2s_{j}x_{j}.
\end{equation}

We now return to the Bell inequality in \eqref{ineq1} and focus on even $N$. We apply $N$ transformations described by the bit-strings $\textbf{s}$: $(0,0,...,0)$ (the all-zeroes bit-string) and the $(N-1)$ bit-strings $\textbf{s}$ that all have $s_{N}=1$, and only one other element being equal to one, e.g. $(1,0,0,...,1)$ or $(0,1,0,...,1)$. All of these bit-strings have an even number of ones, therefore $\sum_{j=1}^{N}s_{j}=2k$ for $k\in\{0,1\}$. For correlators in the inequality of \eqref{ineq1}, the inputs $\textbf{x}$ satisfy $\sum_{j=1}^{N}x_{j}\textrm{ mod } 2=0$, so that $\sum_{j=1}^{N}x_{j}=2k'$ for $k'$ being some integer. Therefore all the correlators $\langle x\rangle$ that appear in \eqref{ineq1} are mapped to $(-1)^{2(k+k'-\sum_{j=1}^{N}s_{j}x_{j})}\langle x\rangle=\langle x\rangle$ and thus the transformation does not alter the inequality.

To obtain $N$ bits of global randomness for even $N$, we choose the input $\textbf{x}'=(1,1,...,1,0)$, the bit-string of all-ones except $x'_N=0$. This input does not appear in \eqref{ineq1} and indeed $\sum_{j=1}^{N}x_{j}=2k'+1$ for some integer $k'$, therefore the above transformations map $\langle x'\rangle$ to $(-1)^{1+2(k+k'-\sum_{j=1}^{N}s_{j}x_{j})}\langle x'\rangle=-\langle x'\rangle$. Due to the uniqueness of the probability distribution maximally violating the Bell inequality, $\langle x'\rangle=-\langle x'\rangle=0$.

We now need to show that all correlators $\langle x'_{J_{n}}\rangle$ where $x'_{J_{n}}$ is the string of $n<N$ elements from $\textbf{x}'=(1,1,...,1,0)$ for a sub-set $J_{n}$. We consider the Hamming distance $H(x'_{J_{n}},s_{J_{n}})$ between $x'_{J_{n}}$ and the corresponding string $s_{J_{n}}$ of elements of $\textbf{s}$ where $s_{j}$ is in $s_{J_{n}}$ if $j\in J_{n}$. Immediately we see that for at least one string $s_{J_{n}}$, the Hamming distance is $H(x'_{J_{n}},s_{J_{n}})=(n-1)$. Therefore, there is at least one transformation $\textbf{s}$ that maps $\langle x'_{J_{n}}\rangle$ to $(-1)^{n-(n-1)}\langle x'_{J_{n}}\rangle=-\langle x'_{J_{n}}\rangle$ for all $J_{n}$. Again, given that all correlators should be equal after the transformation we have that $\langle x'_{J_{n}}\rangle=-\langle x'_{J_{n}}\rangle=0$.

To summarize, we have shown that all correlators that appear in \eqref{distro} for the input $\textbf{x}'=(1,1,...,1,0)$ are equal to zero if the probability distribution that produces them maximally violates the inequality in \eqref{ineq1}. This therefore implies that $p(\textbf{a}|\textbf{x}')=\frac{1}{2^{N}}$ for all $\textbf{a}$ and for all even $N$. For this input $\textbf{x}'$ we obtain $N$ bits of global randomness. We can use another inequality to obtain $N$ bits of global randomness for odd $N$ as shown in Ref. \cite{dhara}. Therefore, we can obtain $N$ bits of global randomness for all $N$. We have used the fact that there is a unique \textit{quantum} violation of the inequality in \eqref{ineq1}. However, for more general theories this may not be the case.

\subsection{Proof of Result 3}

We present a proof that it is possible to certify $3$ bits of global randomness for a set of correlations that is strictly larger than the quantum set $Q$. We call this set $Q^{1+ABC}$ in the terminology of the multipartite generalization of the Navascu\'{e}s-Pironio-Ac\'{i}n hierarchy of correlations that can be characterized through semi-definite programming \cite{npa}. We prove this result utilising the tripartite Mermin inequality \cite{Mermin1990a}, so we are therefore in the $(3,2,2)$ scenario.

We first recall from Ref. \cite{npa} that correlations $p(a_{1},a_{1},a_{3}|x_{1},x_{1},x_{3})$ are contained in the set $Q^{1+ABC}$ if there exists a pure quantum state $|\psi\rangle$, and projectors $\{E_{x_{1}}^{a_{1}},F_{x_{1}}^{a_{1}},G_{x_{3}}^{a_{3}}\}$ labelled by inputs $x_{j}\in\{0,1\}$ and outputs $a_{j}\in\{0,1\}$, such that
\begin{enumerate}
\item \textit{(Hermiticity)} -- $(E_{x_{1}}^{a_{1}})^{\dagger}=E_{x_{1}}^{a_{1}}$, $(F_{x_{1}}^{a_{1}})^{\dagger}=F_{x_{1}}^{a_{1}}$, and $(G_{x_{3}}^{a_{3}})^{\dagger}=G_{x_{3}}^{a_{3}}$ for all $x_{j}$ and $a_{j}$
\item \textit{(Normalization)} -- $\sum_{a_{1}}E_{x_{1}}^{a_{1}}=\mathbb{I}$, $\sum_{a_{1}}F_{x_{1}}^{a_{1}}=\mathbb{I}$, and $\sum_{a_{3}}G_{x_{3}}^{a_{3}}=\mathbb{I}$ for all $x_{j}$
\item \textit{(Orthogonality)} -- $E_{x_{1}}^{a_{1}}E_{x_{1}}^{a'_{1}}=\delta^{a_{1}}_{a'_{1}}E_{x_{1}}^{a_{1}}$, $F_{x_{1}}^{a_{1}}F_{x_{1}}^{a'_{1}}=\delta^{a_{1}}_{a'_{1}}F_{x_{1}}^{a_{1}}$, and $G_{x_{3}}^{a_{3}}G_{x_{3}}^{a'_{3}}=\delta^{a_{3}}_{a'_{3}}G_{x_{3}}^{a_{3}}$ for all $x_{j}$,
\end{enumerate}
such that probabilities are $p(a_{1},a_{1},a_{3}|x_{1},x_{1},x_{3})=\langle\psi|E_{x_{1}}^{a_{1}}F_{x_{1}}^{a_{1}}G_{x_{3}}^{a_{3}}|\psi\rangle$. In addition to these general constraints, linear combinations of these probabilities are elements of a positive semidefinite matrix $\Gamma_{1+ABC}\succeq 0$. We choose a specific positive semidefinite matrix with elements $[\Gamma_{1+ABC}]_{ij}=\langle\psi|\mathcal{O}_{i}^{\dagger}\mathcal{O}_{j}|\psi\rangle$ where $\mathcal{O}_{l}\in\{\mathbb{I},\{A_{i}\},\{B_{j}\},\{C_{k}\},\{A_{i}B_{j}C_{k}\}\}$ for $A_{i}=E_{i}^{0}-E_{i}^{1}$, $B_{j}=F_{j}^{0}-F_{j}^{1}$ and  $C_{k}=G_{k}^{0}-G_{k}^{1}$. Therefore the matrix $\Gamma_{1+ABC}$ is a $15$-by-$15$ matrix with each $\mathcal{O}_{l}$ labelling a row or column. We can now make several observations: $\mathcal{O}_{l}\mathcal{O}_{l}=\mathbb{I}$ for all $\mathcal{O}_{l}$ therefore $(\mathcal{O}_{l})^{\dagger}=\mathcal{O}_{l}$; $\langle x_{1}x_{1}x_{3}\rangle=\langle\psi|A_{x_{1}}B_{x_{1}}C_{x_{3}}|\psi\rangle$; $\langle x_{1}x_{1}\rangle=\langle\psi|A_{x_{1}}B_{x_{1}}|\psi\rangle$; $\langle x_{1}x_{3}\rangle=\langle\psi|A_{x_{1}}C_{x_{3}}|\psi\rangle$; $\langle x_{1}x_{3}\rangle=\langle\psi|B_{x_{1}}C_{x_{3}}|\psi\rangle$; $\langle x_{1}\rangle=\langle\psi|A_{x_{1}}|\psi\rangle$; $\langle x_{1}\rangle=\langle\psi|B_{x_{1}}|\psi\rangle$; and $\langle x_{3}\rangle=\langle\psi|A_{x_{3}}|\psi\rangle$. Here we utilized the notation introduced in the previous section. Finally, the set of quantum correlations $Q$ is a subset of $Q^{1+ABC}$ since the former can be recovered from the latter by imposing more constraints on the projectors. It can also be shown that $Q$ is a strict subset of $Q^{1+ABC}$ for all possible scenarios $(N,M,d)$.

Now that we have defined the set $Q^{1+ABC}$ of correlations that concerns us, we return to the issue of randomness certification. We wish to show that for correlations in this set that maximally violate the tripartite Mermin inequality \cite{Mermin1990a}
\begin{equation}
\langle 001\rangle +\langle 010\rangle +\langle 100\rangle -\langle 111\rangle\leq 2,
\end{equation}
$p(\textbf{a}|\textbf{x}_{0})=\frac{1}{8}$ for all $\textbf{a}$ for a particular input $\textbf{x}_{0}$. We choose this input to be $\textbf{x}_{0}=(0,0,0)$ but it will turn out that we could choose any input $\textbf{x}$ that does not appear in the Mermin inequality. The maximal violation of the Mermin inequality is $4$ and because this violation is achievable with quantum mechanics \cite{Mermin1990a} and $Q\subseteq Q^{1+ABC}$, it is achievable in $Q^{1+ABC}$ also. Therefore, ascertaining the maximal probability $p(\textbf{a}|000)$ compatible with this violation and for correlations in $Q^{1+ABC}$ is an optimization of the form:
\begin{align}
\textrm{maximize }& p(\textbf{a}|000)\nonumber\\
\textrm{subject to }& \langle 001\rangle +\langle 010\rangle +\langle 100\rangle -\langle 111\rangle = 4,\nonumber\\
&p(\textbf{a}|000)\in Q^{1+ABC}.
\end{align}
Given our construction of correlations in $Q^{1+ABC}$, we can rephrase this optimization in terms of a semidefinite program:
\begin{align}
\textrm{maximize }& \frac{1}{2}\textrm{tr}(M\Gamma_{1+ABC})\nonumber\\
\textrm{subject to }& \frac{1}{2}\textrm{tr}(B\Gamma_{1+ABC}) = 4,\nonumber\\
&\Gamma_{1+ABC}\succeq 0,&\nonumber\\
&\frac{1}{2}\textrm{tr}(D_{i}\Gamma_{1+ABC})=0,i\in\{1,2,...,m\},\label{sdp}
\end{align}
where $M$, $B$ and $D_{i}$ are real, symmetric $15$-by-$15$ matrices such that $\frac{1}{2}\textrm{tr}(M\Gamma_{1+ABC})=p(\textbf{a}|000)$ and $\frac{1}{2}\textrm{tr}(B\Gamma_{1+ABC})=\langle 001\rangle +\langle 010\rangle +\langle 100\rangle -\langle 111\rangle$. Due to \eqref{distro}, we can impose the former equality on $\frac{1}{2}\textrm{tr}(M\Gamma_{1+ABC})$. The $m$ matrices $D_{i}$ just impose constraints on elements of $\Gamma_{1+ABC}$ such that they are compatible with $Q^{1+ABC}$.

We now fix the particular representation of $\Gamma_{1+ABC}$ with elements $[\Gamma_{1+ABC}]_{ij}=\langle\psi|\mathcal{O}_{i}\mathcal{O}_{j}|\psi\rangle$ such that for both rows $i$ and columns $j$ we write the ordered vector of operators $\mathcal{O}_{i}$ with $i$ increasing from left to right:
\begin{equation}
(\mathcal{O}_{1},...,\mathcal{O}_{15})=(\mathbb{I},A_{0},A_{1},B_{0},B_{1},C_{0},C_{1},A_{0}B_{1}C_{1},A_{1}B_{0}C_{1},A_{1}B_{1}C_{0},A_{0}B_{0}C_{0},A_{1}B_{0}C_{0},A_{0}B_{1}C_{0},A_{0}B_{0}C_{1},A_{1}B_{1}C_{1}).
\end{equation}
Immediately we observe that the diagonal elements of the matrix $[\Gamma_{1+ABC}]_{ii}=1$ and thus the magnitude of all elements of the matrix $|[\Gamma_{1+ABC}]_{ij}|\leq 1$ are bounded if the matrix is positive semidefinite. For example, given this representation $B=C+C^{\textrm{T}}$ where $C=\bigl(\begin{smallmatrix}
v\\ \tilde{0}
\end{smallmatrix} \bigr)$ for $\tilde{0}$ being a $14$-by-$15$ matrix of zeroes and $w=\left(0,0,...,0,1,1,1,-1\right)$.

There is a unique solution to the problem in \eqref{sdp} if instead of the probability distribution being in $Q^{1+ABC}$ it is constrained to be in $Q$. As mentioned $Q\subseteq Q^{1+ABC}$, so we can write this solution as a matrix of the form $\Gamma_{1+ABC}$, and we call this solution matrix $\Gamma_{\textrm{M}}$ and define it as follows:
\begin{definition} The only solution matrix $\Gamma_{\textrm{M}}$ to \eqref{sdp} that can be realized in quantum theory has elements
\begin{enumerate}
\item $[\Gamma_{\textrm{M}}]_{ij}=1$ if $i=j$, $[\Gamma_{\textrm{M}}]_{ij}\in\{\langle 001\rangle,\langle 010\rangle,\langle 100\rangle\}$ and $[\Gamma_{\textrm{M}}]_{ij}\in\{\langle\psi|\mathcal{P}(\mathcal{P}')^{\dagger}|\psi\rangle,\langle\psi|(\mathcal{P}')^{\dagger}\mathcal{P}|\psi\rangle\}$ where $\mathcal{P}$, $\mathcal{P}'\in\{A_{0}A_{1},B_{0}B_{1},C_{0}C_{1}\}$ and $\mathcal{P}\neq\mathcal{P}'$;
\item $[\Gamma_{\textrm{M}}]_{ij}=-1$ if $[\Gamma_{\textrm{M}}]_{ij}=\langle 111\rangle$ and $[\Gamma_{\textrm{M}}]_{ij}\in\{\langle\psi|\mathcal{P}\mathcal{P}'|\psi\rangle,\langle\psi|\mathcal{P}'\mathcal{P}|\psi\rangle\}$ where $\mathcal{P}$, $\mathcal{P}'\in\{A_{0}A_{1},B_{0}B_{1},C_{0}C_{1}\}$ and $\mathcal{P}\neq\mathcal{P}'$;
\item $[\Gamma_{\textrm{M}}]_{ij}=0$ otherwise.
\end{enumerate}
\end{definition}
We now present the main theorem of this section.
\begin{theorem}
The only possible solution matrix $\Gamma_{1+ABC}$ to the semidefinite program in \eqref{sdp} is $\Gamma_{M}$.
\label{thm1}\end{theorem}
This immediately leads to the following corollary that is relevant for randomness certification. That is, since the solution to the semidefinite program in \eqref{sdp} is the quantum solution we inherit the result of Dhara et al \cite{dhara} that shows that we obtain three random bits if we maximally violate the Mermin inequality \cite{Mermin1990a}. We state this result more formally in the following corollary.
\begin{cor}
The maximal value of the objective function $\frac{1}{2}\textrm{tr}(M\Gamma_{1+ABC})=p(\textbf{a}|000)$ in the semidefinite program \eqref{sdp} is equal to $\frac{1}{8}$ for all $\textbf{a}$.
\end{cor}
\textit{Proof} -- First we observe that, as obtained from the definition of matrix $\Gamma_{\textrm{M}}$, $\langle 0\rangle=0$ for every party's single-body correlator, and equally $\langle 00 \rangle=0$ for all two-body correlators between the three parties, and $\langle 000\rangle=0$. Substituting these values into \eqref{distro}, we then obtain $p(\textbf{a}|000)=\frac{1}{8}$ for all $\textbf{a}$. Since $\Gamma_{\textrm{M}}$ is the only possible solution to \eqref{sdp}, this is the only possible probability distribution over $\textbf{a}$. $\square$
\newline

To prove theorem \ref{thm1} we require two lemmas that will be introduced and proved in the sequel. The first lemma describes the structure of the feasible matrices $\Gamma_{1+ABC}$, i.e. the matrices that satisfy all of the constraints in \eqref{sdp}. The second lemma just says that matrices $\Gamma_{1+ABC}$ of this form are positive semidefinite if and only if they are equal to $\Gamma_{\textrm{M}}$. We now present and prove these lemmas. For simplicity we utilize the notation  for the notation $\langle O\rangle=\langle\psi|O|\psi\rangle$ with $O\in\{A_{j},B_{j},C_{j}|j\in\{0,1\}\}$.

\begin{lemma}
Matrices $\Gamma_{1+ABC}$ are feasible (satisfy all constraints therein) for the semidefinite program \eqref{sdp} if and only if they are of the form:
\begin{equation}
\Gamma_{1+ABC} =
 \begin{pmatrix}
  1 & \textbf{q}_{1} & \textbf{q}_{2} & \textbf{q}_{3} \\
  \textbf{q}_{1}^{T} & \mathbb{W} & \mathbb{X} & \mathbb{Y} \\
  \textbf{q}_{2}^{T} & \mathbb{X}^{T} & \mathbb{D} & \mathbb{O} \\
  \textbf{q}_{3}^{T} & \mathbb{Y}^{T} & \mathbb{O} & \mathbb{D}
 \end{pmatrix},
\end{equation}\label{solnmatrix}
with
\begin{eqnarray}
\textbf{q}_{1}&=&\left(\langle A_{0}\rangle, \langle A_{1}\rangle, \langle B_{0}\rangle, \langle B_{1}\rangle, \langle C_{0}\rangle, \langle C_{1}\rangle\right),\nonumber\\
\textbf{q}_{2}&=&\left(0, 0, 0, 0\right),\nonumber\\
\textbf{q}_{3}&=&\left(1, 1, 1, -1\right),\nonumber\\
\mathbb{W}&=&
 \begin{pmatrix}
  \mathbb{I} & \mathbb{C} & \mathbb{B} \\
  \mathbb{C}^{T} & \mathbb{I} & \mathbb{A} \\
  \mathbb{B}^{T} & \mathbb{A}^{T} & \mathbb{I}
 \end{pmatrix},\nonumber\\
 \mathbb{X}&=&
 \begin{pmatrix}
  -\langle A_{1}\rangle & \langle A_{1}\rangle & \langle A_{1}\rangle & \langle A_{1}\rangle \\
  -\langle A_{0}\rangle & \langle A_{0}\rangle & \langle A_{0}\rangle & \langle A_{0}\rangle \\
  \langle B_{1}\rangle & -\langle B_{1}\rangle & \langle B_{1}\rangle & \langle B_{1}\rangle \\
  \langle B_{0}\rangle & -\langle B_{0}\rangle & \langle B_{0}\rangle & \langle B_{0}\rangle \\
  \langle C_{1}\rangle & \langle C_{1}\rangle & -\langle C_{1}\rangle & \langle C_{1}\rangle \\
  \langle C_{0}\rangle & \langle C_{0}\rangle & -\langle C_{0}\rangle & \langle C_{0}\rangle \\
   \end{pmatrix},\nonumber\\
   \mathbb{Y}&=&
 \begin{pmatrix}
  \langle A_{0}\rangle & \langle A_{0}\rangle & \langle A_{0}\rangle & -\langle A_{0}\rangle \\
  \langle A_{1}\rangle & \langle A_{1}\rangle & \langle A_{1}\rangle & -\langle A_{1}\rangle \\
  \langle B_{0}\rangle & \langle B_{0}\rangle & \langle B_{0}\rangle & -\langle B_{0}\rangle \\
  \langle B_{1}\rangle & \langle B_{1}\rangle & \langle B_{1}\rangle & -\langle B_{1}\rangle \\
  \langle C_{0}\rangle & \langle C_{0}\rangle & \langle C_{0}\rangle & -\langle C_{0}\rangle \\
  \langle C_{1}\rangle & \langle C_{1}\rangle & \langle C_{1}\rangle & -\langle C_{1}\rangle \\
   \end{pmatrix},\nonumber\\
   \mathbb{D}&=&
 \begin{pmatrix}
  1 & 1 & 1 & -1 \\
  1 & 1 & 1 & -1 \\
  1 & 1 & 1 & -1 \\
 -1 & -1 & -1 & 1 \\
   \end{pmatrix}\nonumber,
\end{eqnarray}
$\mathbb{O}$ being a $4$-by-$4$ matrix of all-zeroes, with $\mathbb{A}=\bigl(\begin{smallmatrix}
\langle A_{1}\rangle & \langle A_{0}\rangle \\ \langle A_{0} \rangle&-\langle A_{1}\rangle
\end{smallmatrix} \bigr)$, $\mathbb{B}=\bigl(\begin{smallmatrix}
\langle B_{1}\rangle & \langle B_{0}\rangle \\ \langle B_{0}\rangle&-\langle B_{1}\rangle
\end{smallmatrix} \bigr)$, $\mathbb{C}=\bigl(\begin{smallmatrix}
\langle C_{1}\rangle & \langle C_{0}\rangle \\ \langle C_{0}\rangle&-\langle C_{1}\rangle
\end{smallmatrix} \bigr)$ and $\mathbb{I}=\bigl(\begin{smallmatrix}
1 & 0 \\ 0&1
\end{smallmatrix} \bigr)$.
\end{lemma}

\textit{Proof} -- Vectors $\textbf{q}_{1}$ and $\textbf{q}_{3}$ are trivially obtained if the constraints in \eqref{sdp} are satisfied.

We now use the observation that for all feasible matrices $\Gamma_{1+ABC}$, the elements $[\Gamma_{1+ABC}]_{ij}\in\{\langle 001\rangle, \langle 010\rangle, \langle 100\rangle\}$ are all equal to $1$ and when $[\Gamma_{1+ABC}]_{ij}=\langle 111\rangle$ the element is equal to $-1$. This is due to the fact that this is the only combination of values compatible with maximal violation of the Mermin inequality. This fact implies that $\langle\psi|\mathcal{R}|\psi\rangle=\langle\psi|\psi\rangle$ for $\mathcal{R}\in\{A_{0}B_{0}C_{1},A_{0}B_{1}C_{0},A_{1}B_{0}C_{0}\}$ and $\langle\psi|A_{1}B_{1}C_{1}|\psi\rangle=-\langle\psi|\psi\rangle$ and by normalization,
\begin{align}
A_{0}B_{0}C_{1}|\psi\rangle&=|\psi\rangle,\nonumber\\
A_{0}B_{1}C_{0}|\psi\rangle&=|\psi\rangle,\nonumber\\
A_{1}B_{0}C_{0}|\psi\rangle&=|\psi\rangle,\nonumber\\
A_{1}B_{1}C_{1}|\psi\rangle&=-|\psi\rangle.
\label{ghzcond}\end{align}
This implies that $\langle\psi|\mathcal{P}\mathcal{R}|\psi\rangle=\langle\psi|\mathcal{P}|\psi\rangle$ for $\mathcal{R}\in\{A_{0}B_{0}C_{1},A_{0}B_{1}C_{0},A_{1}B_{0}C_{0}\}$ and $\langle\psi|\mathcal{P}A_{1}B_{1}C_{1}|\psi\rangle=-\langle\psi|\mathcal{P}|\psi\rangle$ where $\mathcal{P}$ is any $\mathcal{O}_{j}$ as described above for $j\in\{1,2,...,15\}$. Utilising this observation we obtain the sub-matrix $\mathbb{D}$ in \eqref{solnmatrix} if $\mathcal{P}$ is equal to any of the $\mathcal{R}$ described above. Also for $\mathcal{P}\in\{A_{i},B_{j},C_{k}\}$ for all $i$, $j$, $k$, we again utilize this observation to obtain $\mathbb{Y}$ and certain elements of $\mathbb{X}$. The elements of $\mathbb{X}$ that are obtained via this observation are those where $\langle\psi|\mathcal{O}_{i}\mathcal{O}_{j}|\psi\rangle=\langle\psi|\mathcal{P}\mathcal{R}|\psi\rangle$ with $P$ and $R$ being as described above.

To obtain the remaining elements of $\mathbb{X}$ that do not satisfy the above condition, we utilize another consequence of the conditions of \eqref{ghzcond}. That is, since $\mathcal{O}_{i}\mathcal{O}_{i}=\mathbb{I}$, any element of $\Gamma_{1+ABC}$ equal to $\langle\psi|\mathcal{S}|\psi\rangle$ for $\mathcal{S}\in\{A_{i}B_{j},A_{i}C_{k},B_{j}C_{k}\}$ is equal to $\pm\langle\psi|\mathcal{S}'|\psi\rangle$ for $\mathcal{S}'\in\{A_{i},B_{j},C_{k}\}$ only if $\mathcal{S}\mathcal{S}'\in\{A_{0}B_{0}C_{1},A_{0}B_{1}C_{0},A_{1}B_{0}C_{0},A_{1}B_{1}C_{1}\}$. The sign in front of $\langle\psi|\mathcal{S}'|\psi\rangle$ is determined by the product $\mathcal{S}\mathcal{S}'$. We also use this observation to obtain matrices $\mathbb{A}$, $\mathbb{B}$ and $\mathbb{C}$.

It remains to be shown how the vector $\textbf{q}_{2}$, the matrix $\mathbb{O}$ and the submatrices $\mathbb{I}$ in $\mathbb{W}$ are obtained. We first observe that $\textbf{q}_{2}=(w,x,y,z)$ where $w=\langle\psi|A_{0}B_{1}C_{1}|\psi\rangle$, $x=\langle\psi|A_{1}B_{0}C_{1}|\psi\rangle$, $y=\langle\psi|A_{1}B_{1}C_{0}|\psi\rangle$, and $z=\langle\psi|A_{0}B_{0}C_{0}|\psi\rangle$. Utilising the relations in \eqref{ghzcond}, we obtain
\begin{equation}
\mathbb{O}=
 \begin{pmatrix}
  w & w & w & -w \\
  x & x & x & -x \\
  y & y & y & -y \\
  z & z & z & -z \\
   \end{pmatrix}\label{matrix1}.
\end{equation}
We now observe that $\mathbb{O}$ can be defined in an equivalent way since $\mathcal{O}_{i}\mathcal{O}_{i}=\mathbb{I}$ for all $\mathcal{O}_{i}$. Using this observation and $\langle\psi|\mathcal{O}_{i}\mathcal{O}_{j}|\psi\rangle=\langle\psi|\mathcal{O}_{j}\mathcal{O}_{i}|\psi\rangle$ for $\mathcal{O}_{i}$, $\mathcal{O}_{j}\in\{A_{i},B_{j},C_{k}\}$ and $\mathcal{O}_{i}\neq\mathcal{O}_{j}$, we obtain
\begin{equation}
\mathbb{O}=
 \begin{pmatrix}
  w & \langle C_{0}C_{1}\rangle & \langle B_{0}B_{1}\rangle & \langle A_{0}A_{1}\rangle \\
  \langle C_{0}C_{1}\rangle & x & \langle A_{0}A_{1}\rangle & \langle B_{0}B_{1}\rangle \\
  \langle B_{0}B_{1}\rangle & \langle A_{0}A_{1}\rangle & y & \langle C_{0}C_{1}\rangle \\
  \langle A_{0}A_{1}\rangle & \langle B_{0}B_{1}\rangle & \langle C_{0}C_{1}\rangle & -z \\
   \end{pmatrix}\label{matrix2},
\end{equation}
where again we are using the notation $\langle\psi|\mathcal{O}_{i}\mathcal{O}_{j}|\psi\rangle=\langle\mathcal{O}_{i}\mathcal{O}_{j}\rangle$ for brevity. Since the matrix in \eqref{matrix1} and \eqref{matrix2} have to be equal to each other, the only possible solution is that $\mathbb{O}$ is a $4$-by-$4$ matrix of zeroes. This also implies that $\textbf{q}_{2}=\left(0,0,0,0\right)$ and $\langle A_{0}A_{1}\rangle=\langle B_{0}B_{1}\rangle=\langle C_{0}C_{1}\rangle=0$, thus completing the matrix $\mathbb{W}$. This also completes our proof. $\square$
\newline

We now present our final lemma that will complete the proof of theorem \ref{thm1}.

\begin{lemma}
The matrix $\Gamma_{1+ABC}$ described by \eqref{solnmatrix} is positive semidefinite if and only if $\Gamma_{1+ABC}=\Gamma_{\textrm{M}}$.
\end{lemma}

\textit{Proof} -- We can use the Schur complement of $\Gamma_{1+ABC}$ in \eqref{solnmatrix} and that $\mathbb{D}=\textbf{q}_{3}^{T}\cdot\textbf{q}_{3}$ and $\mathbb{Y}=\textbf{q}_{1}^{T}\cdot\textbf{q}_{3}$ to show that $\Gamma_{1+ABC}$ is positive semidefinite if and only if
\begin{equation}
\begin{pmatrix}
  \mathbb{W}' & \mathbb{X} \\
  \mathbb{X}^{T} & \mathbb{D} \\
\end{pmatrix}\label{submatrix}
 \succeq 0,
\end{equation}
where $\mathbb{W}'=\mathbb{W}-\textbf{q}_{1}^{T}\cdot\textbf{q}_{1}$. For example, for the matrix $\Gamma_{\textrm{M}}$, the corresponding submatrix $\Gamma_{\textrm{M}}'$ from \eqref{submatrix} is
\begin{equation}
\Gamma_{\textrm{M}}'=
\begin{pmatrix}
  \mathbb{I} & \bar{0} \\
  \bar{0}^{T} & \mathbb{D} \\
\end{pmatrix},
\end{equation}
where $\mathbb{I}$ is the $6$-by-$6$ identity matrix and $\bar{0}$ is a $6$-by-$4$ matrix of zeroes. This submatrix of $\Gamma_{\textrm{M}}$ is positive semidefinite if and only if $\mathbb{D}\succeq 0$ which is indeed true.

Since the space of positive semi-definite matrices is convex, the set of feasible matrices $\Gamma_{1+ABC}$ for the semidefinite program \eqref{sdp} is a convex set. Therefore, if there is a submatrix $\Gamma_{1}$ of the form \eqref{submatrix}, we can obtain another submatrix $\Gamma_{2}$ of the form \eqref{submatrix} that is a convex combination of $\Gamma_{1}$ and $\Gamma_{\textrm{M}}'$. We assume that $\Gamma_{1}$ has elements corresponding to some non-zero values $\{\langle A_{i}\rangle, \langle B_{j}\rangle, \langle C_{k}\rangle\}$, therefore completely unlike $\Gamma_{\textrm{M}}$. We now show that there exist matrices of the form $\Gamma_{2}$ that are not positive semidefinite which implies that any matrix $\Gamma_{1}$ as described is not positive semidefinite. This in turn implies that the only positive semidefinite matrix of the form \eqref{submatrix} is $\Gamma_{\textrm{M}}'$.

We choose $\Gamma_{2}$ such that $\sum_{j=0}^{1}|\langle A_{j}\rangle|+|\langle B_{j}\rangle|+|\langle C_{j}\rangle|\ll 1$ but at least one of the elements of the set $\{\langle A_{i}\rangle, \langle B_{j}\rangle, \langle C_{k}\rangle\}$ is non-zero. As mentioned before, since the space of solution matrices $\Gamma_{1+ABC}$ is convex we can always choose such a matrix without loss of generality. Therefore, the matrix in \eqref{submatrix} is positive semidefinite if and only if
\begin{equation}
\begin{pmatrix}
  \bar{\mathbb{W}'} & \bar{\mathbb{X}} \\
  \bar{\mathbb{X}}^{T} & \mathbb{D} \\
\end{pmatrix}
-
\frac{1}{(1-\langle A_{0}\rangle^{2})}\begin{pmatrix}
  \textbf{s}^{T} \\
  \textbf{r}^{T} \\
\end{pmatrix}\cdot\begin{pmatrix}
  \textbf{s} & \textbf{r} \\
\end{pmatrix}\label{newmatrix}
 \succeq 0,
\end{equation}
where $\mathbb{X}=\bigl(\begin{smallmatrix}
\textbf{r} \\ \bar{\mathbb{X}}
\end{smallmatrix} \bigr)$ where $\textbf{r}=\left(-\langle A_{1}\rangle,\langle A_{1}\rangle,\langle A_{1}\rangle,\langle A_{1}\rangle\right)$ is the first row of $\mathbb{X}$, $\bar{\mathbb{W}'}$ is $\mathbb{W}'$ without the first column and first row, and $\textbf{s}$ is the first row of $\mathbb{W}'$ excluding the element $[\mathbb{W}']_{11}$. Since every diagonal element of $\bar{\mathbb{W}'}-\frac{1}{(1-\langle A_{0}\rangle^{2})}\textbf{s}^{T}\cdot\textbf{s}$ is positive by construction, then the matrix in \eqref{newmatrix} is positive semidefinite if and only if $\mathbb{E}=\mathbb{D}-\frac{1}{(1-\langle A_{0}\rangle^{2})}\textbf{r}^{T}\cdot\textbf{r}\succeq 0$. Note that every diagonal element of $\mathbb{E}$ is equal to $1-\frac{1}{(1-\langle A_{0}\rangle^{2})}\langle A_{1}\rangle^{2}$. However, the element $[\mathbb{E}]_{12}=1+\frac{1}{(1-\langle A_{0}\rangle^{2})}\langle A_{1}\rangle^{2}$. For a matrix to be positive semedefinite off-diagonal elements have a magnitude that is bounded by the diagonal terms, therefore for $\mathbb{E}$ to be positive semidefinite we must satisfy $\langle A_{1}\rangle=0$.

We can now repeatedly apply the same analysis to subsequent bottom-left submatrices of \eqref{newmatrix} where the matrix in \eqref{submatrix} is positive semidefinite if and only if $\mathbb{E}'=\mathbb{D}-\frac{1}{\alpha}\textbf{r'}^{T}\cdot\textbf{r'}\succeq 0$ where $\alpha<1$ is some positive real number and $\textbf{r'}$ is any row of $\mathbb{X}$. For every matrix $\mathbb{E}'$ the diagonal elements are $1-\frac{1}{\alpha}\langle \mathcal{P}\rangle^{2}$ where $\mathcal{P}\in\{A_{i},B_{j},C_{k}\}$ but there are off-diagonal terms in $\mathbb{E}'$ that take the value $1+\frac{1}{\alpha}\langle \mathcal{P}\rangle^{2}$. Therefore for all $\Gamma_{1+ABC}$ described by \eqref{solnmatrix}, $\langle \mathcal{P}\rangle=0$ for $\mathcal{P}\in\{A_{i},B_{j},C_{k}\}$ for all $i$, $j$, $k$. This matrix thus corresponds to $\Gamma_{\textrm{M}}$ and completes our proof. $\square$
\newline

Combining the two lemmas above we then obtain our proof of Theorem \ref{thm1}. This concludes our observation that maximally random numbers can be certified within a set of correlations that is not the quantum set. Our proof is analytic and makes concrete the numerical observations in Dhara et al \cite{dhara}. It would be interesting to extend this proof to other scenarios even though we have used a lot of the structure of the Mermin inequality and the $(3,2,2)$ scenario.

\end{document}